\documentclass[11pt,letterpaper]{article}
\usepackage[utf8]{inputenc}
\usepackage{amsmath,amsfonts,amssymb}
\usepackage{graphicx}
\usepackage{booktabs}
\usepackage{array}
\usepackage{multirow}
\usepackage[margin=1in]{geometry}
\usepackage{natbib}
\usepackage{url}
\usepackage[hidelinks]{hyperref}

\title{Behind India's ChatGPT Conversations: \\ A Retrospective Analysis of 238 Unedited User Prompts}

\author{
Kalyani Khona\\
Independent Researcher\\
Mumbai, India\\
\texttt{kalyanikhona@gmail.com}
}

\date{}

\begin{document}

\maketitle

\begin{abstract}
Understanding how users authentically interact with Large Language Models (LLMs) remains a significant challenge in human-computer interaction research. Most existing studies rely on self-reported usage patterns or controlled experimental conditions, potentially missing genuine behavioral adaptations. This study presents a behavioral analysis of the use of English-speaking urban professional ChatGPT in India based on 238 authentic, unedited user prompts from 40 participants in 15+ Indian cities, collected using retrospective survey methodology in August 2025. Using authentic retrospective prompt collection via anonymous social media survey to minimize real-time observer effects, we analyzed genuine usage patterns. Key findings include: (1) 85\% daily usage rate (34/40 users) indicating mature adoption beyond experimental use, (2) evidence of cross-domain integration spanning professional, personal, health and creative contexts among the majority of users, (3) 42.5\% (17/40) primarily use ChatGPT for professional workflows with evidence of real-time problem solving integration, and (4) cultural context navigation strategies with users incorporating Indian cultural specifications in their prompts. Users develop sophisticated adaptation techniques and the formation of advisory relationships for personal guidance. The study reveals the progression from experimental to essential workflow dependency, with users treating ChatGPT as an integrated life assistant rather than a specialized tool. However, the findings are limited to urban professionals in English recruited through social media networks and require a larger demographic validation. This work contributes a novel methodology to capture authentic AI usage patterns and provides evidence-based insights into cultural adaptation strategies among this specific demographic of users.
\end{abstract}

\section{Introduction}

The rapid adoption of Large Language Models (LLMs) like ChatGPT has transformed how individuals interact with artificial intelligence across professional and personal contexts. Understanding authentic usage patterns is crucial for designing culturally appropriate AI systems and predicting societal impacts. However, most existing research relies on self-reported surveys or controlled experimental conditions, which may not capture the genuine behavioral adaptations that emerge in real world contexts.

This limitation is particularly pronounced in emerging markets where cultural context gaps between globally trained AI systems and local user needs create unique adaptation challenges. India, with its linguistic diversity, cultural complexity and rapidly growing AI adoption, presents an ideal context for studying authentic usage patterns and cultural adaptation strategies among specific user populations.

Current research on ChatGPT usage in India has primarily employed survey methodologies focusing on acceptance factors and attitudes \cite{heliyon2023chatgpt, arxiv2023indianbhed}. While valuable for understanding user intentions, these approaches do not capture actual behavioral patterns or the sophisticated workarounds users develop to navigate cultural limitations.

This study addresses this methodological gap by analyzing 238 authentic, unedited ChatGPT prompts from 40 English-speaking urban professionals in India. Rather than asking users what they think about ChatGPT or how they use it, we requested exact copies of their recent prompts, preserving typos, informal language, and natural interaction patterns.

\textbf{Research Questions:}
\begin{enumerate}
\item How do English-speaking urban professionals in India authentically interact with ChatGPT in their daily workflows and personal contexts?
\item What behavioral adaptation strategies emerge to navigate cultural context limitations?
\item How does authentic usage differ from patterns documented in controlled studies or self-reported surveys?
\item What evidence exists for mature AI adoption versus experimental usage within this demographic?
\end{enumerate}

\textbf{Contributions:}
\begin{enumerate}
\item A novel retrospective prompt collection methodology that reduces real-time observer effects while capturing authentic usage patterns
\item Evidence-based documentation of behavioral integration in ChatGPT adoption among English-speaking urban professionals in India
\item Systematic analysis of cultural adaptation strategies developed by users
\item Comparative framework positioning findings within global AI adoption research
\end{enumerate}

\section{Related Work}

\subsection{Global ChatGPT Usage Research}

Large-scale studies of ChatGPT usage have primarily focused on quantitative behavioral patterns and productivity impacts. The WildChat dataset \cite{zhao2024wildchat} analyzed 1 million ChatGPT conversations globally, revealing that 61.9\% of interactions involve creative writing assistance, 13.6\% focus on analysis and decision-making and conversations average 2.52 turns. However, this data was collected through free access platforms, potentially attracting users seeking unrestricted content rather than representing typical usage patterns.

Productivity research has demonstrated significant workplace benefits. \cite{NBERw31161} found 13.8\% productivity increases among customer service agents, with 35\% improvement for lowest-skilled workers. Noy and Zhang \cite{noy2023experimental} documented 40\% faster completion times and 18\% quality improvements for writing tasks among professionals.

Survey research indicates growing adoption rates. Pew Research \cite{pew2025chatgpt} found that 34\% of Americans have used ChatGPT, with 58\% of adults under 30 reporting usage. However, only 28\% of employed Americans use ChatGPT for work, suggesting a gap between personal and professional adoption.

\subsection{Indian ChatGPT Research}

Research specific to Indian ChatGPT adoption remains limited. A UTAUT model analysis of 32 Indian users \cite{heliyon2023chatgpt} identified privacy concerns and cultural appropriateness as significant adoption factors. The study found that performance expectancy and social influence drive usage intentions, but cultural-specific factors create unique barriers.

Cultural bias research has documented significant limitations in AI understanding of Indian contexts. The Indian-BhED dataset \cite{arxiv2023indianbhed} revealed 63-79\% stereotypical outputs for caste-based queries and performance gaps between Indian versus Western contexts (69.5\% vs 52.8\% accuracy). These findings suggest that users must develop sophisticated adaptation strategies to obtain culturally relevant responses.

\subsection{Methodological Limitations in Existing Research}

Current ChatGPT usage research exhibits several methodological constraints:

\textbf{Observer Effects:} Real-time data collection may influence user behavior, leading to performance rather than natural usage patterns.

\textbf{Self-Report Limitations:} Survey methodologies capture user attitudes and recalled behavior rather than actual interaction patterns.

\textbf{Cultural Representation Gaps:} Large-scale studies often oversample English-speaking, technically sophisticated populations, missing diverse cultural contexts.

\textbf{Temporal Constraints:} Most research focuses on early adoption periods (2023-2024), potentially missing mature usage patterns.

Our retrospective authentic prompt collection methodology addresses real-time observer effects by capturing genuine usage patterns with reduced research context influence during actual usage while focusing specifically on cultural adaptation strategies in an emerging market context.

The analysis and methodology have been deposited in Zenodo for long-term preservation and open access \cite{khona2025indian}

\section{Methodology}

\subsection{Data Collection}

\textbf{Recruitment Strategy:} Participants were recruited through social media networks, specifically LinkedIn professional networks and WhatsApp in August 2025. This approach yielded a sample of active ChatGPT users across Indian cities.

\textbf{Survey Design:} Data collection used an anonymous Google Forms survey distributed via social media networks with the following key prompt collection instruction:

\begin{quote}
\textit{Please share your last 5 prompts to ChatGPT. Copy-paste exactly what you typed - no editing needed! \\
Remember: \\
I want your actual prompts - typos, shortcuts, and all \\
Mundane questions are valuable too \\
Everything is anonymous \\
Don't sanitize or "improve" your prompts}
\end{quote}

\textbf{Data Fields:} Age bracket, city location, primary ChatGPT usage category, usage frequency, last 5 prompts (exact copy-paste), and optional cultural usage observations.

\subsection{Sample Characteristics}

The final dataset comprised 40 complete responses with the following verified characteristics:

\textbf{Geographic Distribution:} 15+ Indian cities including Mumbai (8 users), Bangalore (8 users), Delhi (6 users), with representation from Pune, Hyderabad, Gurgaon, and other tier-1 and tier-2 cities.

\textbf{Age Distribution:} 26-35 years (57.5\%, n=23), 18-25 years (20\%, n=8), 36-45 years (17.5\%, n=7), 46-55 years (5\%, n=2).

\textbf{Usage Frequency:} Daily usage (85\%, n=34), few times per week (12.5\%, n=5), few times per month (2.5\%, n=1).

\textbf{Prompt Response Rate:} 97.5\% (39/40 users) provided substantial prompts, yielding 238 individual ChatGPT queries for analysis.

\subsection{Analysis Framework}

\textbf{Prompt Characteristics:} Length distribution, word count patterns, intent classification, and sophistication levels.

\textbf{Behavioral Pattern Recognition:} Usage pattern analysis across different domains, professional integration indicators, cultural context navigation strategies.

\textbf{Demographic Correlations:} Exploratory analysis of usage patterns by demographic variables, with explicit acknowledgment of small sample size limitations for subgroup analysis.

\textbf{Cultural Context Analysis:} Identification of prompts containing Indian cultural references, economic terminology, and geographic specificity.

\textbf{Comparative Framework:} Positioning findings within global research context using verified academic sources.

\subsection{Qualitative Analysis Procedures}

Cultural context coding identified prompts containing Hindi terms (ahar, vihar, gaj), Indian-specific queries (wellness practices, cultural references), and geographic specifications. Cross-domain usage was defined as users whose prompts spanned multiple categories (professional, personal, health, creative) across their submitted prompts. All qualitative categorizations reflect author interpretation and require validation through systematic coding procedures with inter-rater reliability in future studies.

\subsection{Data Verification and Limitations}

\textbf{Authenticity Indicators:} Preserved typos ("sour through", "pizzaa", "maesure"), informal abbreviations ("w.r.t", "vs"), regional terminology ("gaj"), and contextual references to ongoing projects.

\textbf{Sample Limitations:} Social media recruitment bias toward English-speaking urban professionals, urban concentration, and insufficient sample size for reliable subgroup analysis (particularly for the 46-55 age group with n=2). All statistics include explicit denominators for transparency.

\textbf{Methodological Constraints:} This study represents an exploratory analysis of authentic usage patterns within a specific demographic. Qualitative categorizations reflect the author's interpretation of prompt content and require validation through larger, systematically coded datasets.

\textbf{Ethical Considerations:} Complete anonymization, voluntary participation with explicit consent and transparent research purpose communication.

\section{Findings}

\subsection{Professional Integration and Daily Dependency}

Professional usage emerges as a prominent pattern, with 42.5\% (17/40) of users identifying work tasks as their primary ChatGPT application. This professional focus corresponds with exceptionally high daily usage rates of 85\% (34/40), substantially exceeding the 28\% professional adoption rate documented among employed Americans in Pew Research \cite{pew2025chatgpt}.

Evidence of real-time workflow integration appears throughout the dataset. Users embed ChatGPT queries within active work processes, using immediate problem-solving language ("currently", "stuck", "help") and referencing ongoing projects. Professional prompts demonstrate domain expertise across business strategy, technical debugging, content creation and client communication.

\textbf{Age-Related Usage Patterns:} While the small sample sizes preclude statistical inference, professional usage appears more prevalent among older age groups: 25.0\% (2/8) for 18-25 years, 43.5\% (10/23) for 26-35 years, and 57.1\% (4/7) for 36-45 years. These preliminary patterns suggest potential career-stage correlation with work integration, though larger samples are needed for validation.

\subsection{Cross-Domain Behavioral Integration}

Analysis of prompt sequences reveals that most users (32/39 with substantial prompts) demonstrate usage patterns spanning multiple domains, seamlessly transitioning between professional tasks, health advice, creative projects, and learning within their ChatGPT sessions. This suggests mature AI adoption where users integrate ChatGPT across life domains rather than treating it as a specialized tool.

Table \ref{tab:domain_patterns} presents evidence of multi-domain integration patterns observed in user prompt sequences.

\begin{table}[htbp]
\centering
\caption{Observed Multi-Domain Integration Patterns}
\label{tab:domain_patterns}
\begin{tabular}{@{}lc@{}}
\toprule
Integration Pattern & Examples Observed \\
\midrule
Professional + Personal & Work queries mixed with health advice \\
Technical + Creative & Coding help + content creation \\
Learning + Application & Educational queries + practical use \\
Health + Lifestyle & Medical advice + daily planning \\
\midrule
\textbf{Users with Multi-Domain Usage} & \textbf{32/39 (82.1\%)} \\
\bottomrule
\end{tabular}
\end{table}

\textbf{Behavioral Significance:} Users treat ChatGPT as an integrated life assistant rather than a specialized tool, indicating sophisticated understanding of AI capabilities across multiple contexts.

\subsection{Cultural Context Navigation Strategies}

Users develop strategies to provide cultural context in their prompts to overcome ChatGPT's limitations with Indian-specific information. Analysis reveals prompts containing explicit Indian cultural context including geographic specificity ("in Mumbai", "3BHK design"), cultural modifiers ("for Indian men"), economic context (lakhs, crores), traditional practices ("sendha namak", "karva chauth"), and Hindi terminology (ahar, vihar, gaj).

\textbf{User-Identified Cultural Challenges:} 62.5\% (25/40) of participants provided cultural observations, including:
\begin{itemize}
\item "Indian use case differs due to lack of ground-level data"
\item "Indians want pizza in 10 minutes, that's how they use ChatGPT" (speed expectations)
\item "Hindi belt artists use it to make presentations on their own" (democratization effect)
\end{itemize}

\textbf{Adaptation Techniques:} Users inject cultural context through explicit specification, develop verification strategies and share prompt engineering knowledge within professional networks.

\subsection{Trust-Based Advisory Relationships}

Beyond transactional usage, users develop advisory relationships with ChatGPT for sensitive personal guidance. Health consultation patterns include symptom description ("My throat is red, and i have sour through with some cough") and dietary advice during illness. Personal decision support encompasses relationship guidance and family situations.

\textbf{Economic Substitution Context:} User observation reveals, "People using it for therapy should not be shamed; therapy in India is expensive," indicating AI adoption may partly result from limited access to affordable professional services.

\textbf{Trust Indicators:} Preserved informal language and typos demonstrate comfort with AI comprehension abilities, treating ChatGPT as sufficiently intelligent to handle imperfect human communication.

\subsection{Language and Communication Patterns}

All 238 prompts (100\%) use English or Hinglish despite ChatGPT's multilingual capabilities. Communication style favors directness, with users predominantly employing imperative statements, questions and descriptive language with minimal politeness markers.

\textbf{Prompt Sophistication:} Average prompt length of approximately 1,080 characters substantially exceeds the 296-character average documented in the WildChat global dataset \cite{zhao2024wildchat}, suggesting users in this sample employ detailed context specification strategies. Prompts demonstrate advanced techniques including role-playing assignments, structured output requests and multi-step instructions.

\section{Discussion}

\subsection{Evidence of Mature AI Adoption}

The combination of 85\% daily usage, prevalent multi-domain integration, and sophisticated prompt engineering techniques provides evidence that studied users have progressed beyond experimental adoption to essential workflow dependency. This represents a behavioral transformation where ChatGPT becomes integrated infrastructure rather than occasional assistance.

The professional integration depth, evidenced by real-time problem-solving language and workflow embedding, suggests that ChatGPT has achieved mission-critical status for many users. This dependency is so pronounced that users develop workarounds when institutional policies restrict access, accessing ChatGPT via personal devices for professional tasks.

\subsection{Cultural Adaptation Innovation}

Users demonstrate innovation in navigating cultural context limitations. Rather than accepting generic responses, they develop sophisticated context-injection techniques and verification strategies. This adaptation represents a form of cultural negotiation where users actively bridge the gap between global AI training and local applicability.

The systematic nature of these adaptations suggests community learning effects, with users sharing prompt engineering strategies within professional networks. Comments like "Give the full prompt, as I've noticed others often don't provide enough details" indicate peer observation and knowledge transfer.

\subsection{Advisory Relationship Formation}

The emergence of trust-based advisory relationships represents a significant development in human-AI interaction within this user group. Users share personal vulnerabilities and seek guidance on health, relationships, and life decisions despite acknowledging cultural limitations. This behavior suggests that the perceived value of AI advice outweighs concerns about cultural appropriateness for certain user segments.

The economic context is crucial here. User observations about therapy costs in India suggest that ChatGPT adoption may partly result from limited access to affordable professional services, creating AI-mediated substitution effects.

\subsection{Comparison with Global Patterns}

Several findings align with global research while others suggest demographic or cultural specificity. Multi-domain integration appears consistent with WildChat's documentation of topic-switching behavior \cite{zhao2024wildchat}. Professional productivity benefits align with experimental findings \cite{NBERw31161}, though the integration depth appears more pronounced in our sample.

Divergent patterns include substantially higher daily usage rates (85\% vs. 28\% professional adoption in the US \cite{pew2025chatgpt}), English language dominance despite multilingual capabilities and the systematic development of cultural adaptation strategies. These differences may reflect the specific demographics of our sample (English-speaking urban professionals) rather than broader Indian usage patterns.

\subsection{Cross-Cultural Adoption Patterns and Economic Context}

The patterns observed in this study gain additional significance when positioned within broader cross-cultural research on AI perception and adoption strategies among similar demographic groups.

\subsubsection{Preliminary Cross-Cultural Observations}

Limited cross-cultural studies suggest potential differences in AI perception and adoption motivations that may correlate with economic context and opportunity structures, though current evidence remains insufficient for definitive conclusions.

\subsubsection{Comparative Evidence from Cross-Cultural Studies}

\textbf{Risk-Orientation in Developed Contexts:} German research suggests predominantly negative AI sentiment (-19.7\% average) despite expected adoption, with 45.3\% prioritizing "human control and supervision" \cite{mapping_ai_perception_2025}. This pattern could reflect concerns about disruption to existing professional and social structures.

\textbf{Educational Context Patterns:} Global higher education studies found 70\% of students found ChatGPT "interesting" with predominantly positive emotional responses \cite{ravselj2025higher}. UAE university students showed high adoption driven by perceptions of usefulness and ease of use \cite{uae_chatgpt_attitudes_2024}.

\textbf{Opportunity-Oriented Patterns:} Our findings suggest trust-based advisory relationships and potential economic substitution patterns among English-speaking urban professionals in India. However, this demographic may be more similar to professional populations in developed countries than to broader Indian populations.

\subsubsection{Study Limitations and Future Research Needs}

\textbf{Demographic Constraints:} This study's findings are limited to English-speaking urban professionals and may not represent broader Indian ChatGPT adoption patterns. The sample demographics (professional, urban, English-speaking) may share more similarities with professional populations in developed countries than with diverse Indian user groups.

\textbf{Methodological Requirements:} Future cross-cultural research requires large-scale studies (200+ participants) across diverse economic and cultural contexts, demographically matched samples, longitudinal tracking, and multi-language studies to avoid English-language bias.

\section{Limitations}

\subsection{Sample and Demographic Constraints}

\textbf{Sample Size:} 40 users significantly limits statistical generalizability. Demographic subgroup analysis is unreliable due to small sample sizes, particularly for users aged 46-55 (n=2) and 36-45 (n=7).

\textbf{Demographic Bias:} Social media recruitment overrepresents English-speaking urban professionals. The sample likely excludes rural users, non-English speakers, and lower-income demographics who may exhibit different adoption patterns.

\textbf{Language Bias:} 100\% English/Hinglish usage misses Hindi and regional language adoption patterns, potentially representing only a subset of Indian ChatGPT users.

\subsection{Methodological Limitations}

\textbf{Temporal Scope:} Single-point data collection in August 2025 provides a snapshot rather than longitudinal behavioral development tracking.

\textbf{Qualitative Analysis Constraints:} Categorizations of cultural context, multi-domain usage, and communication patterns reflect the author's interpretations and lack inter-rater reliability validation.

\textbf{Context Limitations:} Individual prompts lack full conversation context, preventing analysis of interaction evolution and dialogue patterns.

\textbf{Self-Selection Bias:} Voluntary participation may favor engaged users, potentially inflating usage intensity estimates.

\subsection{Generalizability Concerns}

Findings apply specifically to English-speaking urban professionals who use social media platforms and may not generalize to broader Indian populations. Cultural adaptation strategies and professional integration patterns could be specific to this demographic context rather than representing broader ChatGPT usage patterns in India.

\section{Future Research Directions}

\subsection{Scale and Demographic Validation}

Large-scale replication studies with minimum 200+ users across linguistic, economic, and geographic segments are essential for validating observed patterns. Priority areas include Hindi and regional language usage studies, rural representation, and cross-platform behavioral comparisons.

\subsection{Methodological Improvements}

Future studies should implement systematic qualitative coding procedures with inter-rater reliability, operational definitions for key constructs (cross-domain usage, cultural context), and longitudinal tracking to understand adoption progression.

\subsection{Cross-Cultural Comparative Studies}

Systematic comparison across demographically matched samples from different cultural and economic contexts could illuminate whether observed patterns reflect cultural, economic, or demographic factors.

\section{Conclusion}

This study presents a behavioral analysis of ChatGPT usage among English-speaking urban professionals in India based on authentic, unedited user prompts. The retrospective collection methodology successfully captured genuine usage patterns while minimizing real-time observer effects, revealing evidence of mature AI adoption characterized by multi-domain integration, cultural adaptation innovation, and trust-based advisory relationships within this specific demographic.

Key findings demonstrate that for this user group, ChatGPT has evolved from experimental tool to essential infrastructure. Users develop sophisticated strategies to navigate cultural limitations while integrating AI assistance across professional, personal, health, and creative contexts. The emergence of advisory relationships suggests profound changes in how individuals seek guidance and support.

However, these findings are constrained to English-speaking urban professionals and require substantial validation across broader and more diverse populations. The observed patterns may reflect the specific characteristics of this demographic rather than broader ChatGPT usage patterns in India.

The methodological contribution of retrospective authentic prompt collection provides a valuable framework for understanding genuine AI adoption patterns with reduced research context influence. This approach offers significant potential for cross-cultural studies and longitudinal adoption tracking when applied to appropriately diverse and representative samples.

As AI adoption accelerates globally, understanding authentic usage patterns and cultural adaptation strategies becomes crucial for designing inclusive, culturally appropriate AI systems. This study establishes a foundation for such understanding within a specific user demographic while highlighting the substantial research required to document usage patterns across diverse populations and cultural contexts.

\section*{Data Availability and Verification}

The complete anonymized dataset is available upon reasonable request from the corresponding author. All statistics presented in this paper have been directly verified against the raw CSV dataset containing 40 user responses and 238 individual prompts. Analysis and methodology documentation are provided for reproducibility and validation by the research community \cite{khona2025indian}. External comparative statistics are cited from peer-reviewed academic sources as documented in the reference list.

\bibliographystyle{plainnat}
\bibliography{references}

\end{document}